\documentclass[aps,prd,showpacs,nobibnotes,twocolumn,preprintnumbers,
nobalancelastpage,superscriptaddress]{revtex4}
\usepackage{latexsym}
\usepackage{dcolumn}
\usepackage{bm}
\usepackage{amsmath}
\usepackage{natbib}
\input{epsf}
\newcommand{\be}{\begin{equation}}
\newcommand{\ee}{\end{equation}}
\newcommand{\ba}{\begin{eqnarray}}
\newcommand{\ea}{\end{eqnarray}}
\newcommand{\bi}{\begin{itemize}}
\newcommand{\ei}{\end{itemize}}
\newcommand{\bfi}{\begin{figure}[!t]
\epsfxsize=9cm
\epsffile}
\newcommand{\bfig}{\begin{figure*}[t]
\epsfxsize=15cm
\epsffile}
\newcommand{\efi}{\end{figure}}
\newcommand{\efig}{\end{figure*}}
\newcommand{\no}{\nonumber}
\newcommand{\mpch}{{\rm Mpc}/h}
\newcommand{\hmpc}{h/{\rm Mpc}}
\newcommand{\la}{\lesssim}
\newcommand{\ga}{\gtrsim}

\begin{document}
\title{Sampling artifact in volume weighted velocity measurement.--- II. Detection in simulations and comparison with theoretical modelling}
\author{Yi Zheng}
\affiliation{Key Laboratory for Research in Galaxies and Cosmology, Shanghai
  Astronomical Observatory, 80 Nandan Road, Shanghai 200030,
  China}
\affiliation{Korea Astronomy and Space Science Institute, Daejeon 305-348, Republic of
Korea}
\author{Pengjie Zhang}
\email[Email me at: ]{zhangpj@sjtu.edu.cn}
\affiliation{Center for Astronomy and Astrophysics, Department of
  Physics and Astronomy, Shanghai Jiao Tong
  University, 955 Jianchuan road, Shanghai 200240, China}
\affiliation{Key Laboratory for Research in Galaxies and Cosmology, Shanghai
  Astronomical Observatory, 80 Nandan Road, Shanghai 200030,
  China}
\author{Yipeng Jing}
\email[Email me at: ]{ypjing@sjtu.edu.cn}
\affiliation{Center for Astronomy and Astrophysics, Department of
  Physics and Astronomy, Shanghai Jiao Tong
  University, 955 Jianchuan Road, Shanghai 200240, China}


\begin{abstract}
Measuring the volume weighted velocity power spectrum suffers from a
severe systematic error due to  imperfect sampling of the velocity
field from the inhomogeneous distribution of dark matter particles/halos
in simulations or galaxies with velocity measurement. This ``sampling
artifact'' depends on both the mean particle number density $\bar{n}_P$
and the intrinsic large scale structure (LSS) fluctuation in the particle distribution.   (1) We report  robust detection of
this sampling artifact in N-body simulations. It causes $\sim 12\%$
underestimation of the velocity power  spectrum at $k=0.1\hmpc$ for
samples with $\bar{n}_P=6\times  10^{-3}
(\mpch)^{-3}$. This systematic underestimation increases with
decreasing $\bar{n}_P$ and increasing $k$. Its dependence on the
intrinsic LSS fluctuations is also robustly detected. (2) All of these findings are expected based upon our theoretical modelling
in paper I \cite{Zhang14}.  In particular, the leading order
theoretical approximation agrees quantitatively well with the simulation
result for $\bar{n}_P\ga 6\times 10^{-4} (\mpch)^{-3}$. Furthermore, we
provide an ansatz to take high order terms into account. It improves
the model accuracy to $\la 1\%$ at $k\la 0.1\hmpc$ over 3 orders of
magnitude in $\bar{n}_P$ and over typical LSS
clustering from $z=0$ to $z=2$. (3) The sampling 
artifact is determined by the deflection ${\bf D}$ field, which is
straightforwardly available in both simulations and data of galaxy
velocity. Hence the sampling artifact in the velocity power
spectrum measurement can be self-calibrated within our framework. By
applying such self-calibration in simulations,  it is promising to determine the
{\it real} large scale velocity bias of $10^{13}M_\odot$ halos with
$\sim 1\%$  accuracy, and that of lower mass halos with better
accuracy. (4) In contrast to suppressing the velocity power
spectrum at large scale, the sampling artifact causes an
overestimation of the velocity dispersion. We prove that 
correlation between the signal field (${\bf v}$) and the sampling
field (${\bf D}$) is a major cause. This complexity, among others, shall be carefully
investigated to further improve understanding of the sampling artifact. 
\end{abstract}
\maketitle

\section{Introduction}
\label{sec:intro}
Peculiar velocity is a powerful probe of cosmology, with increasing
importance. A statistics of particular importance to peculiar velocity
cosmology is the {\it volume weighted} velocity power
spectrum. Unlike the density weighted velocity, it is free of
uncertainties in galaxy density bias, which is hard to predict from
first principle.  So the volume weighted velocity is desired for the
purpose of cosmology.  However it is challenging to measure it accurately, both in 
simulations and in observations with galaxy velocity measurement. The
measurement suffers from the
sampling artifact \cite{DTFE96}, which
arises from the fact that we often cannot fairly sample the {\it
  volume weighted} velocity field. For example, the distribution of galaxies with
velocity measurement through distance indicators
(e.g. \cite{Watkins14,6df2014}) is not only sparse 
but also spatially clustered. Even worse, their spatial distribution
is correlated with the velocity field that we try to
measure, due to the underlying correlation between the large scale
structure (LSS) and velocity. Hence the sampling of  
volume weighted velocity field is {\it biased}. 

This sampling artifact  has three-fold impacts on cosmology.  (1) The
velocity power spectrum (and higher order statistics) measured through galaxy
velocity data is systematically biased by this sampling artifact. (2) The 
same sampling artifact also exists in measuring the velocity power
spectrum of dark matter (DM) particles/halos in N-body
simulations. This can systematically bias our theoretical
understanding of the velocity field, for example the physical velocity
bias of halos. (3) A biased theoretical
understanding can lead to biased cosmological
constraints,  even if the velocity measurements themselves, such as
that inferred from redshift space distortion,  are free of the sampling
artifact.  Hence this sampling artifact is  
entangled in key ingredients of peculiar velocity cosmology. It is a
significant source of systematic errors, which we should  investigate
intensively. Throughout this paper, we will focus on its impact on
peculiar velocity power spectrum. Unless otherwise 
specified, we always refer to the peculiar velocity power spectrum as the
volume weighted one.

 In \cite{Zhang14} (hereafter paper I)  we present a theoretical modelling of the sampling
artifact in measuring the volume weighted velocity power spectrum. The
present paper (paper II)
focuses on its numerical verification  in order to improve our
understanding of the sampling artifact. In \cite{Zheng14b} we will apply
this improved understanding to correct the sampling artifact in the
halo/galaxy velocity field and robustly measure the halo velocity
bias.  

In paper I we find that the sampling artifact is fully captured by the
``deflection'' field ${\bf D}$. ${\bf D}$  is the spatial separation 
vector pointing from a particle used for velocity assignment to a grid point
that the velocity is assigned with this particle. Within this
framework, we predict that the sampling artifact causes
underestimation in the velocity power spectrum at large
scale. Furthermore, this systematic underestimation
increases with decreasing particle number density $\bar{n}_P$ and increasing
$k$. With a number of simplifications we are able to derive analytical
expressions for this underestimation. We estimate that it is
significant,  $\sim 10$\% at $k=0.1\hmpc$, for 
$\bar{n}_P=10^{-3}(\mpch)^{-3}$. Without correcting it, the velocity
bias of $10^{13}M_\odot$ halos measured in N-body simulations will be systematically
underestimated by $\sim 5\%$, 
from its real value.  This systematic underestimation/error is larger
than the expected statistical error in peculiar velocity 
determination from redshift space distortion (RSD) by surveys like
BigBOSS/MS-DESI \cite{BigBOSS}, Euclid
\footnote{http://sci.esa.int/euclid/} and SKA
\footnote{https://www.skatelescope.org/}. Furthermore, it is of
comparable size and sign as the {\it physical} velocity bias ($b_v<1$) predicted
through proto-halo statistics
\cite{BBKS,DesjacquesI,DesjacquesII}. Hence it  could mislead the
theory comparison, if not corrected.   Therefore the sampling artifact
is clearly a severe obstacle to theoretical understanding and
observational application of peculiar velocity.  

 The existence of sampling artifact in the volume weighted
  velocity power spectrum has been realized in
  simulations \cite{paperII,Jennings14,Zheng14b}, for both the DM and
  halo velocity fields.  These works found that its impact increases
  with decreasing number  density. This behavior can be used to
  diagnose the sampling artifact \cite{paperII,Jennings14}.  For
  example, by comparing the velocity power spectrum $P_v$ of two halo
  populations at equal volume  density,   difference in $P_v$ is
  significantly suppressed \cite{Jennings14}, manifesting the
  existence of sampling artifact.  One complexity is that the sampling artifact
  also depends on the spatial clustering of halos/DM particles
  \cite{Zhang14}.  Hence two halo populations
  of identical number density but different mass do not have identical sampling artifact,
  due to difference in their spatial clustering. This complexity makes clean separation of
  sampling artifact from a possible halo velocity bias highly nontrivial,
  since the halo number density,  density bias and velocity bias can
  all vary with the halo mass.  In contrast,   since all randomly selected DM populations at
the same redshift have statiatically identical spatial clustering,  the sampling
artifact can be robustly isolated, as done in \cite{paperII}.  The present paper
extends its numerical verification  and quantification to much wider
range of DM density and redshift, and is better theory motivated and
backed up \cite{Zhang14}.  By comparing with the sampling artifact of
DM populations with various number density at various redshifts, we hope to
develop  generic  theoretical modelling of the sampling artifact,
applicable to a wide range of particle number density and spatial
clustering. We can then
robustly predict the sampling artifact in halo velocity and circumvent
the difficulty of directly measuring it.  

Our paper is organized as follows. In \S \ref{sec:detection} we
report the detection of sampling artifact in the volume weighted
velocity power spectrum measurement, including its dependence
on $\bar{n}_P$ and redshift/spatical clustering. In \S
\ref{sec:comparison} we compare it with the theoretical modelling
developed in paper I \cite{Zhang14}.  We also propose an ansatz to further improve
its accuracy. In \S \ref{sec:conclusion} we discuss the possibility to self-calibrate this
sampling artifact .  The appendix \S
\ref{sec:NA} \& \S \ref{sec:v2} discuss more  aspects of the sampling
artifact,  other than the suppression of power discussed in the main
text. These aspects are important and deserve further
investigation.

\section{Detection of the sampling artifact in simulations}
\label{sec:detection}
For brevity, we will focus on the gradient part of the velocity ${\bf
  v}_E$ ($\nabla\times {\bf v}_E=0$), which contains most of
cosmological information. Given a sample of simulation particles/halos/galaxies
with velocity information, we can measure  the volume weighted velocity power
spectrum $\hat{P}_E(k)$. The hat $\hat{}$ denotes the measured quantity, instead of the
one without measurement error (the sampling artifact to be specific). The measurement  can be done using
one's favorite velocity assignment method, such as the ones based on
Voronoi and Delaunay 
tessellations \cite{DTFE96}. Throughout this paper, we restrict to the
NP (Nearest Particle) method \cite{paperII}.  As discussed in paper I,
sampling artifacts in other velocity assignment methods are similar. So
results on the sampling artifact in the NP method also provide 
useful reference for that in other methods. 

\subsection{The method to detect the sampling artifact}
Without knowing the {\it correct} velocity power spectrum $P_E(k)$, we are not
able to carry out direct comparison with $\hat{P}_E(k)$ to measure the
sampling artifact. This problem is circumvented in \cite{paperII}.  We randomly select a
fraction $f$ of simulation DM (dark matter) particles to construct a sub-sample. We then apply the same analysis to this sub-sample to
measure the velocity power spectrum, which we denote as
$\hat{P}_E(k|f)$. So the measurement using the whole 
sample is $\hat{P}_E(k|f=1)$. If there is no sampling artifact, we should
have $\hat{P}_E(k|f)=\hat{P}_E(k|f=1)$, since simulation particles in the sub-sample are
selected randomly from the full sample without prejudice \footnote{An example is to
  measure the matter power spectrum. Choosing a fraction of particles
  only affects shot noise. So as long as we subtract shot noise
  correctly, the power spectra measured using different $f$ of DM
  samples drawing from the same simulation are statistically
  identical.  }. Hence the ratio
\be
\label{eqn:eta}
\eta(k|f)\equiv \frac{\hat{P}_E(k|f)}{\hat{P}_E(k|f=1)}
\ee
measures the sampling artifact \footnote{Strictly speaking, $\eta$
  measures the {\it relative} sampling artifact. It is relative in the
sense that even the full sample ($f=1$) suffers from the sampling
artifact due to its finite number density.  The {\it absolute}
sampling artifact has to be compared with a population with infinite
number density ($f\rightarrow \infty$). But for the purpose of
detecting and understanding the sampling artifact, this relative
measure is sufficient. Furthermore, at least for DM velocity, routine
N-body simulations with $\bar{n}_P\ga 1(\mpch)^{-3}$ and above are
essentially free of sampling artifact if the full sample ($f=1$) is
used. So $\eta$ measured for DM velocity is essentially a measure of
absolute sampling artifact. }. In another word, if $\eta\neq 1$, the
sampling artifact exists. 
\bfi{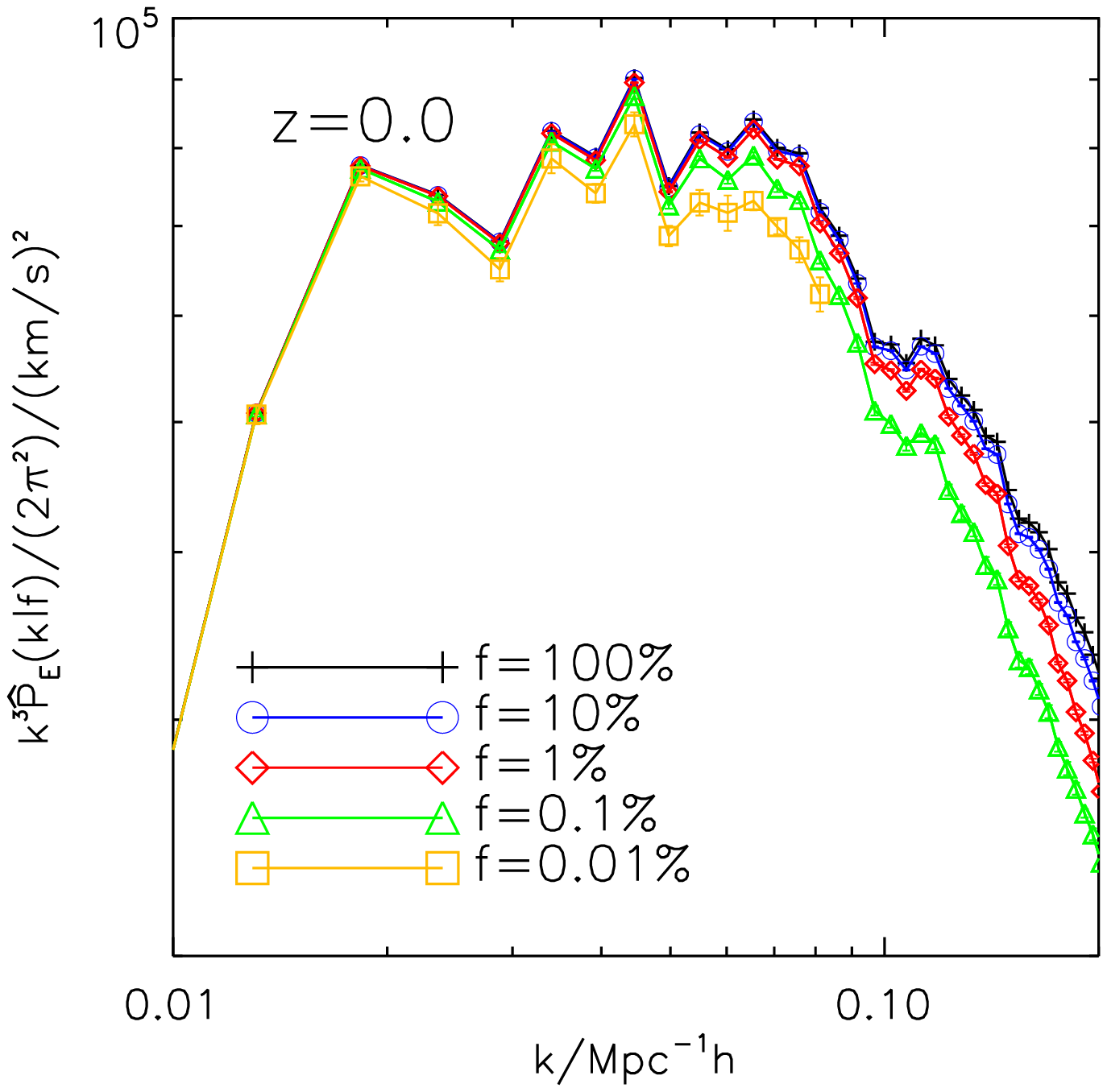}
\caption{The measured velocity power spectra of DM samples at $z=0$
  with various number density. To highlight the
  sampling artifact, we randomly select a fraction $f$ of DM particles
to construct DM sub-samples and then measure the corresponding
velocity power spectrum. Without the sampling artifact, the measured power
spectrum $\hat{P}_E(k|f)$ should be identical to that of the full
sample $\hat{P}_E(k|f=100\%)$. However, in simulations we find a
systematic suppression, which increases with decreasing number density
and increasing $k$. The error bars of sub-samples are estimated
using $10$ sub-samples of identical $f$. For the J1200 simulation
specification, the mean particle number density
$\bar{n}_P=0.62f(\mpch)^{-3}$. \label{fig:vps}} 
\efi

\bfi{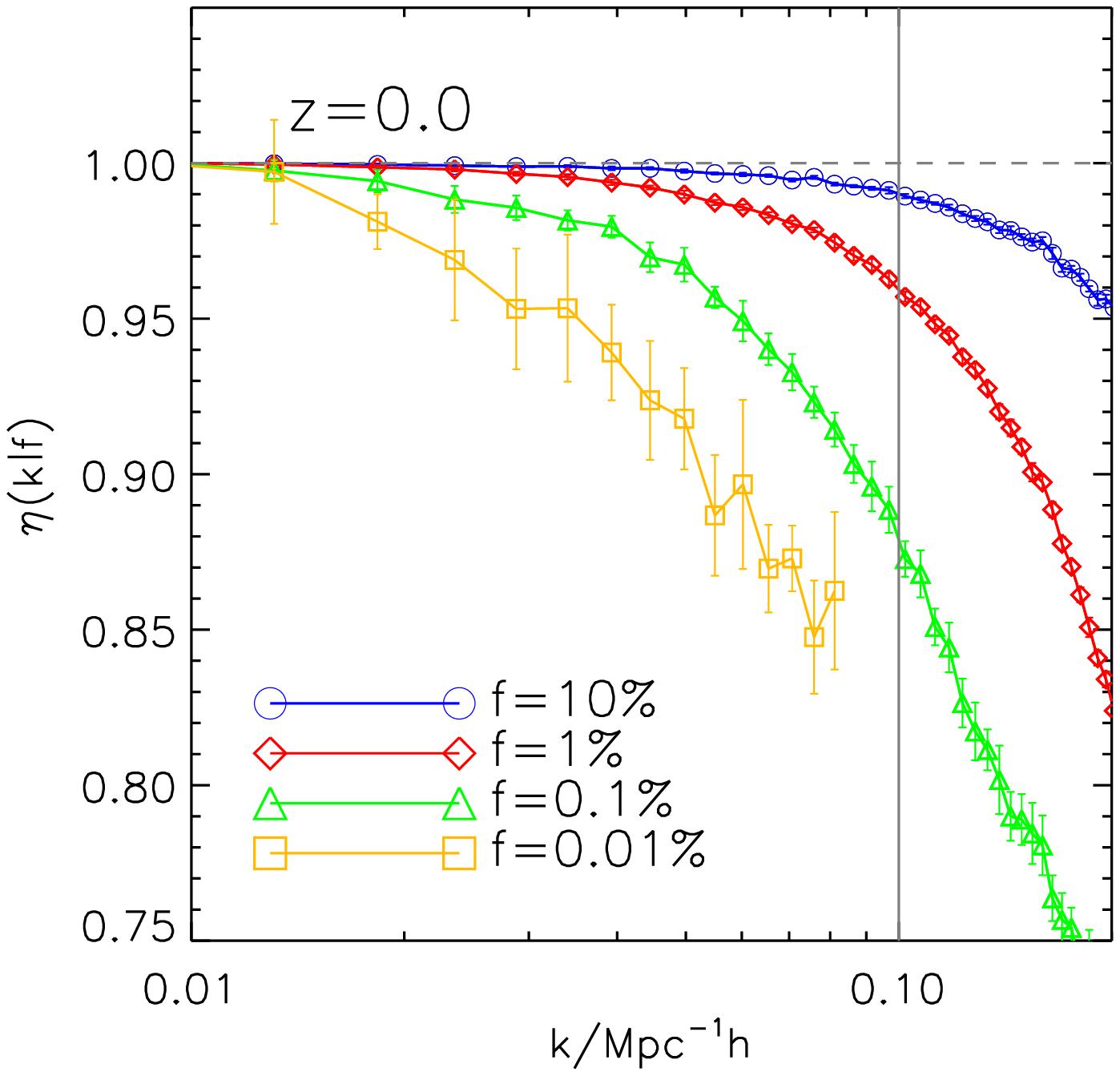}
\caption{$\eta(k|f)\equiv \hat{P}_E(k|f)/\hat{P}_E(k|f=1)$ at
  $z=0$. $\eta<1$ means systematic  underestimation of  the velocity
  power spectrum, caused by the sampling artifact. This systematic
  underestimation can be severe, even at relatively large scale
  $k=0.1\hmpc$,  increasing from $1\%$ for $f=10\%$) 
  ($\bar{n}_P=6.2\times 10^{-2}(\mpch)^{-3}$) to $\sim 12\%$ $f=0.1\%$ 
  ($\bar{n}_P=6.2\times 10^{-4}(\mpch)^{-3}$). \label{fig:eta0}} 
\efi

This method of measuring sampling artifact can be applied to both DM
particles and halos. But due to low number density of halos, the
measurement of $\eta$ is noisy. So we will  focus on $\eta(k|f)$ of DM
particles. Nevertheless, we expect the results
to be general, not limited to the case of DM particles, for two
reasons. (1) As addressed in paper I, the sampling artifact is
determined by the deflection field ${\bf D}$, which is determined both by
$\bar{n}_P$ and the intrinsic  LSS fluctuation in the particle
distribution. By analyzing DM sub-samples with different $f$ at
different simulation snapshots, we cover not only a large parameter
space in $\bar{n}_P$, but also different intrinsic LSS clustering.  (2) We also  use these results on DM particles
to test and improve our theoretical  understanding. Our  theoretical
modelling does not make assumption on whether the sample dealt with
is DM particles or DM halos. Hence we expect that, as long as the
theoretical modelling works for DM particles, it should work for
halos as well. 

\bfi{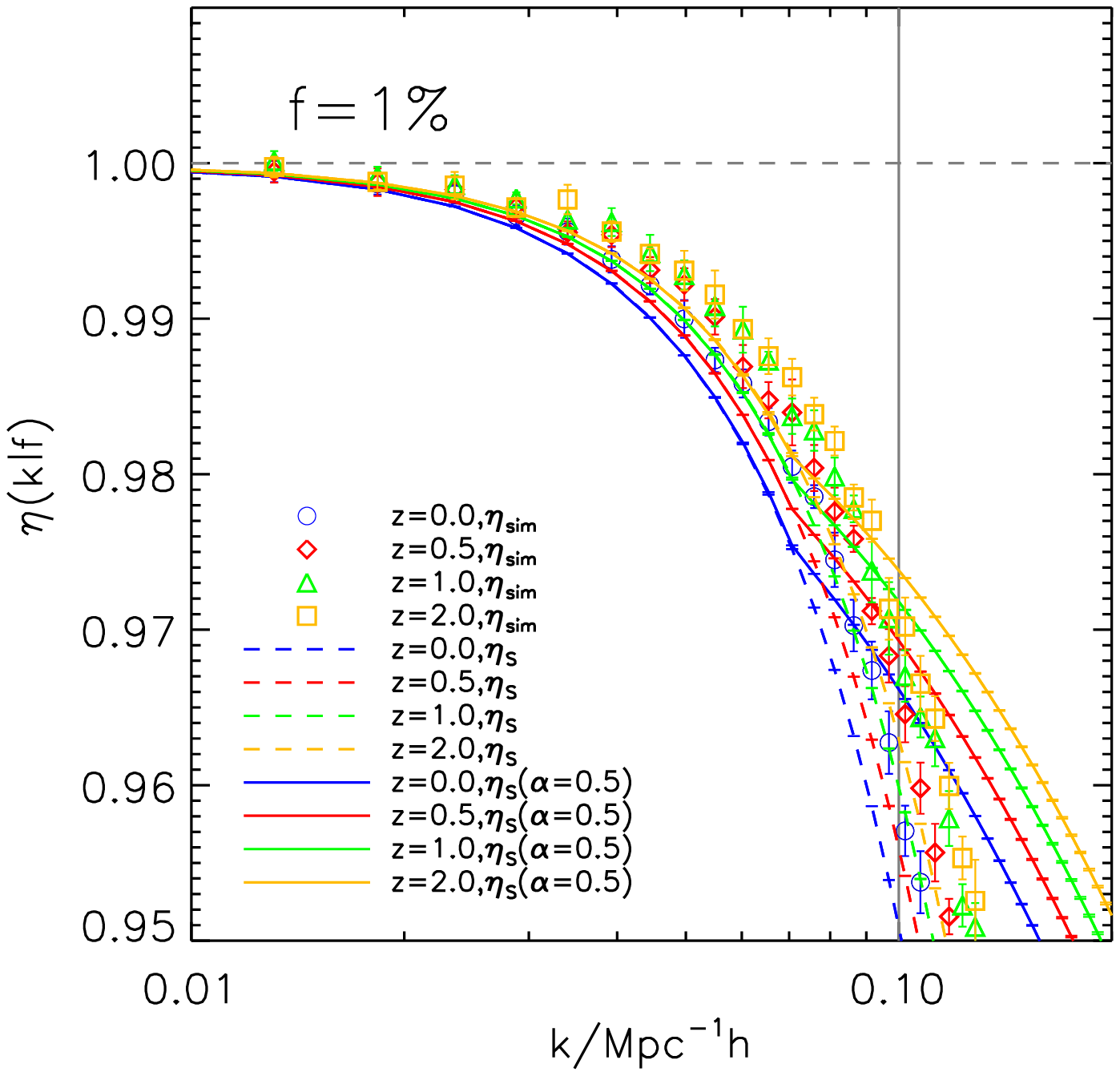}
\caption{$\eta(k|f=1\%)$ at $z=0.0$, $0.5$, $1.0$ and $2.0$, in
  simulations (data points) and in theory (curves). $f=1\%$
  corresponds to $\bar{n}_P=6.2\times 10^{-3} (\mpch)^{-3}$. $\eta$
  increases with increasing redshift.  This
  redshift dependence reflects that the sampling artifact depends not
  only on $\bar{n}_P$, but also the intrinsic LSS clustering. The dash curves are the
  theoretical predictions by paper I (Eq. \ref{eqn:LSA}). The solid
  lines are that of the improved ansatz Eq. \ref{eqn:ISA}, which takes the
  spatial correlation of the ${\bf D}$ field into account. The
  irregularities at $k\sim 0.07\hmpc$ is caused  by  interpolation
  of limited data points of measured ${\bf D}$ correlation, not a  fundamental feature. 
 \label{fig:eta1} }
\efi
\subsection{The sampling artifact and its dependence on the mean
  number density}
We analyze the same J1200 N-body simulation used in \cite{paperII}. It adopts the $\Lambda$CDM cosmology with $\Omega_m=0.268$,
$\Omega_\Lambda=0.732$, $\sigma_8=0.85$,  $n_s=1$ and $h=0.71$. It has
$1024^3$ simulation DM particles and boxsize of $1200 \mpch$. It was run
with a particle-particle-particle-mesh (${\rm P^3M}$) code
\cite{Jing07}. More simulation
details are presented in \cite{paperII}.  For a sub-sample of DM
particles with fraction $f$, the corresponding particle number density
is $\bar{n}_P=0.62f (\mpch)^{-3}$. We use $N_{\rm grid}=256^3$ to
analyze the velocity field and the deflection ${\bf D}$ field. 

Paper I predicts  $\eta<1$ at large scale. In \cite{paperII} we have
already found $\eta<1$, 
for $f=10\%$ ($\bar{n}_P= 0.062 (\mpch)^{-3}$). The current paper will examine the sampling artifact for wider
range of number density ($6.2\times 10^{-5}$-$0.62 (\mpch)^{-3}$),
covering that of $10^{12}$-$10^{13} M_\odot$ halos at $z\in[0,2]$. 

Fig. 1 shows $\hat{P}_E(k|f)$ with $f=100\%$, $10\%$, $1\%$, $0.1\%$ and $0.01\%$ at
$z=0$ and Fig. 2 shows $\eta(k|f)$.  In the velocity power spectrum
measurement we have subtracted shot noise following
\cite{Zhang14}. The alias effect \cite{Jing05,Pueblas09,Koda13,Zhang14}
still exists. But the alias effect does not vary with $f$. So $\eta$
isolates  the sampling artifact.

\bfi{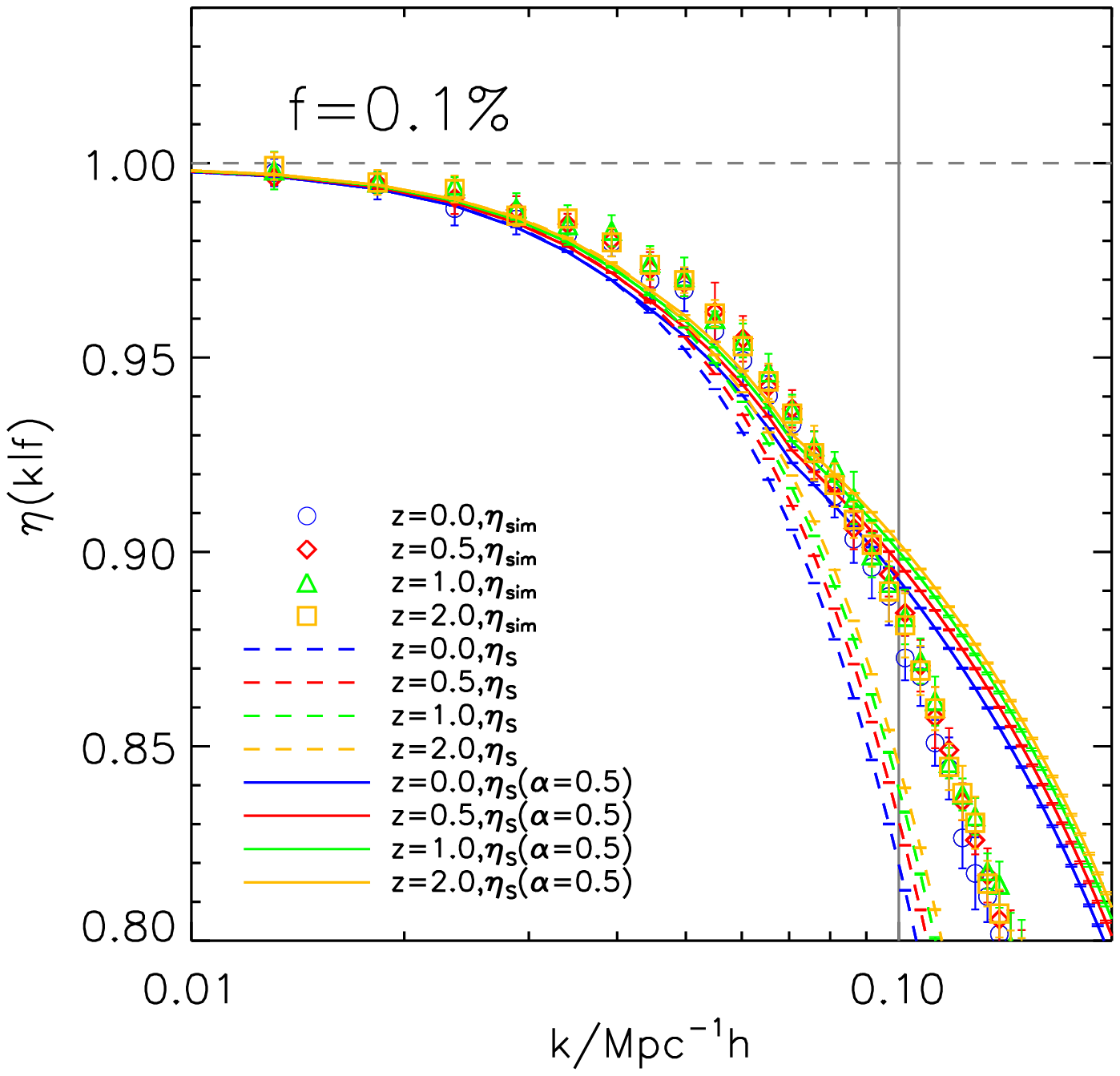}
\caption{Same as Fig. \ref{fig:eta1}, but for $f=0.1\%$ ($\bar{n}_P=6.2\times
  10^{-4} (\mpch)^{-3}$). Eq. \ref{eqn:ISA} (solid curves) improves over
  Eq. \ref{eqn:LSA} (dashed curves) from $\sim 6\%$ at
  $k=0.1\hmpc$ to $1\%$. \label{fig:eta01} }
\efi

We detect the sampling artifact at high significance. (1)
Fig. \ref{fig:vps} \& \ref{fig:eta0} clearly show systematic underestimation ($\eta<1$) of $P_E$, which should not exist without the 
sampling artifact. The result of $f=10\%$ confirms our previous finding in \cite{paperII}. 
(2) The underestimation 
increases with decreasing $f$ ($\bar{n}_P$).  For $f=0.1\%$ ($\bar{n}_P=6.2\times
10^{-4}(\mpch)^{-3}$), $\eta=0.88$ at $k=0.1\hmpc$. This number density
corresponds to $\sim 10^{13}M_\odot$ halos at $z=0$. This means
that the velocity power spectrum of $10^{13}M_\odot$ halos measured
without correcting the sampling artifact
can be wrong by $\sim 10\%$, leading to a systematic error of $\delta
b_v\sim -0.05$ in the halo velocity bias measurement. This is certainly a
significant source of systematic error to be worried about and
investigated heavily. (3) The systematic underestimation/error increases with increasing
$k$. Therefore it is more challenging to understand the sampling
artifact and infer cosmology at smaller scales.  

\bfi{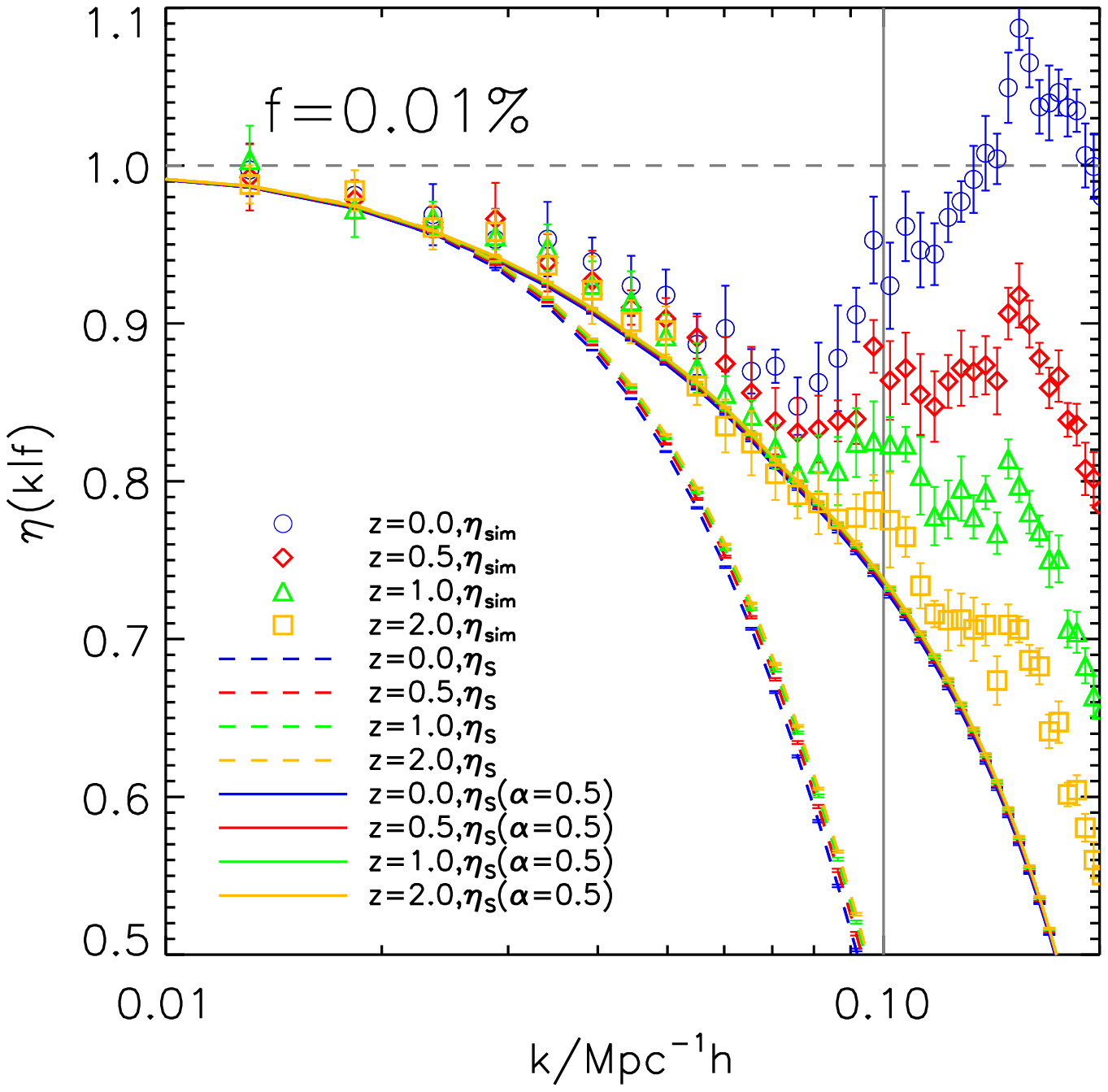}
\caption{The same as Fig. \ref{fig:eta1} \& \ref{fig:eta01}, but for
  $f=0.01\%$ ($\bar{n}_P=6.2\times 10^{-5}
  (\mpch)^{-3}$). Eq. \ref{eqn:ISA} significantly improves over
  Eq. \ref{eqn:LSA}).  Data points 
  at $k\ga 0.08\hmpc$ show  anomalous behaviors. These behaviors can
  not be described within robust treatment of the ${\bf v}$-${\bf D}$
  correlation, neglected so far in the theoretical modelling.  We will
  discuss these complexities in the appendix \S \ref{sec:NA}, show the
  existence of the ${\bf v}$-${\bf D}$ correlation  in  \S
  \ref{sec:vD}, and  its impact in \S \ref{sec:v2} .   \label{fig:eta001}} 
\efi

\subsection{The sampling artifact depends on the intrinsic clustering}
Fig. 3, 4 \& 5 show $\eta$ at redshift $z=0.0,0.5,1.0,2.0$, with
$f=1\%$, $0.1\%$ and $0.01\%$ respectively. For $f=1\%$, the variation
with $z$ is significant. For a fixed
$f$, the DM samples at different $z$ only differ in their  intrinsic LSS
fluctuation. From $z=0$
to $z=2$, the clustering amplitude decreases by a factor $\sim 2.4$ in
linear regime and larger factors in nonlinear regime. Hence the
dependence on $z$ must be caused by 
the evolution in the intrinsic clustering. Therefore this redshift
dependence proves that, besides $\bar{n}_P$,  the intrinsic LSS
fluctuation also
affects the sampling  artifact. 

However,  for $f=0.1\%$, the 
redshift dependence is already insignificant (Fig. 4). These behaviors can be interpreted as
competition between two sources affecting the ${\bf D}$ field, 
namely Poisson fluctuation and intrinsic LSS fluctuation in the particle
distribution. The former is determined by $\bar{n}_P$. The latter decreases towards higher redshift. The two factors both contribute and amplify the underestimation
($\eta<1$). Larger $f$
(e.g. $f=1\%$) means smaller Poisson fluctuation and hence more
significant impact of LSS and redshift dependence \footnote{For
$f=0.01\%$, we found opposite dependence on redshift  (Fig. 5) to the
cases of $f=1\%$ and $0.1\%$. But given the
irregularities in the data and the abnormal increase at $k\ga
0.1\hmpc$, we suspect other impacts of the sampling artifact, which
will be briefly discussed in \S 
\ref{sec:NA}.}.  This point will be elaborated later in \S
\ref{sec:comparison}.  

A brief summary of this section is that we have robustly detected the
sampling artifact. We further identify two factors affecting the
sampling artifact,  $\bar{n}_P$ and the intrinsic LSS fluctuation. It is now the question whether the theoretical
modelling can well reproduce these findings.

\section{Testing and improving the  theoretical modelling}
\label{sec:comparison}
We now proceed to comparison between the theory and
simulation,  to  quantify the accuracy of our model and to improve it. The ultimate
goal is to develop an accurate method to correct for the sampling artifact. It can be
used for two purposes. First is to accurately measure the halo
velocity power spectrum and velocity bias in simulations, with the sampling artifact
corrected. Such measurements at $1\%$ accuracy are needed to compare
with the velocity power spectrum determined indirectly from RSD to
infer the nature of dark matter, dark energy 
and gravity. Second, it can be applied to galaxy velocity data such as
SFI++ \cite{Watkins14} and
6dF \cite{6df2014} to measure the sampling artifact corrected velocity power
spectrum. 

\subsection{Theoretical modelling of the sampling artifact}
Here we briefly summarize our theoretical modelling of the sampling
artifact in paper I \cite{Zhang14}.  It targets at the NP velocity assignment
method \cite{paperII}, but it can also be extended to methods based on various
tessellation methods. In the NP method we approximate the velocity
on a given grid point at 
position ${\bf x}$ as that of the nearest simulation
particle/halo/galaxy at position ${\bf x}_P({\bf x})$,
\be
\label{eqn:hatv}
\hat{{\bf v}}({\bf x})={\bf v}({\bf x}_P({\bf x}))\ .
\ee
Hence the sampling artifact is fully captured by  the  ``deflection'' field
\be
{\bf D}({\bf x})\equiv {\bf x}_P({\bf x})-{\bf x}\ .
\ee 
The sampling artifact arises from ${\bf D}\neq {\bf 0}$. This
distinguishes from other numerical artifacts such as the alias effect
in measuring the velocity power spectrum \cite{Pueblas09,Koda13}.

The velocity power spectrum measured on uniform grids, after subtracting shot noise, is
\ba
\label{eqn:vv2}
\hat{P}_{ij}({\bf k})=\sum_{\bf q}P_{ij}({\bf q}) W({\bf q},{\bf q}^{'})\ .
\ea
Here, $\hat{P}_{ij}({\bf k})\propto \langle \hat{v}_i({\bf
  k})\hat{v}^{*}_j({\bf k})\rangle$. ${\bf k}$ and ${\bf
  q}$ are discrete Fourier modes\footnote{But ${\bf k}$ is bounded, while
${\bf q}$ is not. For a cubic volume with size $L$ and grids $N_{\rm grid}$, ${\bf
  k}=2\pi/L(i,j,k)$ with $|i|\leq N_{\rm grid}^{1/3}/2$. Refer to
paper I for more details. }.  ${\bf q}^{'}\equiv {\bf q}-{\bf k}$.  The window function $W$ is
\be
\label{eqn:W}
W({\bf q},{\bf q}^{'})\equiv \frac{1}{N^2_{\rm grid}} \sum_{{\bf x}\neq{\bf x}^{'}} S({\bf q},{\bf r}) e^{i {\bf q}^{'}\cdot {\bf r}}\ .
\ee
Here ${\bf r}\equiv {\bf x}^{'}-{\bf x}$. $N_{\rm grid}$ is the total number
of grid points. The window function $W({\bf q},{\bf q}^{'})$ is {\it
  inhomogeneous} since it depends not only on ${\bf q}^{'}\equiv {\bf
  q}-{\bf k}$, but also on ${\bf q}$. It makes the deconvolution  to
obtain the true velocity power  spectrum more difficult. 

$W({\bf q},{\bf q}^{'})$ is the Fourier
transform of the sampling function $S({\bf q},{\bf r})$ over ${\bf
  r}$. Imperfect sampling causes $S\neq 1$ and hence results in the
sampling artifact. Under reasonable approximation (however refer to
the appendix \S \ref{sec:NA} \& \ref{sec:v2} for caveat) we obtain
\be
\label{eqn:S}
S({\bf q},{\bf r})=\left\langle e^{i{\bf q}\cdot ({\bf D}^{'}-{\bf D})} \right\rangle\ .
\ee
The ${\bf D}$ field is known in simulations or surveys with galaxy
velocity measurement. Hence $S$ and $W$ can both be
calculated. In the limit that the alias effect can be neglected, namely
now ${\bf k}$ and ${\bf q}$ occupy the same space,  in principle we
can solve Eq. \ref{eqn:vv2} to obtain the true velocity power
spectrum. Unfortunately numerical evaluation of $W({\bf q},{\bf q}^{'})$ is time
consuming. So far we are able to reduce the calculation of all $({\bf
  q},{\bf q}^{'})$ pairs from brute-force computation of size $O(N^3_{\rm grid})$
to $O(N^2_{\rm grid})$ (Eq. 27, paper I).  But further reduction in
computation is still needed to solve Eq. \ref{eqn:vv2} for the true velocity power
spectrum. In paper I and the current paper, we take approximations to
simplify Eq. \ref{eqn:vv2} for efficient evaluation of the sampling
artifact.

${\bf D}\neq {\bf 0}$ leads to $S({\bf q},{\bf r}) <1$. A generic
prediction is that the sampling artifact causes underestimation in the 
velocity power spectrum at large scale \footnote{This statement is valid as long as
  the power of velocity correlation dominates over the power transported
  by spatial correlation in ${\bf D}$. The power of velocity
  correlation concentrates at large scale (e.g. $k\sim 0.1\hmpc$,
  Fig. 1). ${\bf D}$ correlates  at scales $\la L_P$, and redistribute power in ${\bf v}$ over such
  scale. Hence as long as $k\la 1/L_P$, we expect underestimation in
  the velocity power spectrum.   }.  In the limit of no spatial correlation in
${\bf D}$, we are able to derive the leading order sampling artifact (Eq. 36, paper I),
\be 
\label{eqn:LSA}
\hat{P}^{(1)}_E(k)\simeq P_E(k) S(k)\ .
\ee
Here, 
\be
\label{eqn:SKP}
S(k)\equiv S({\bf k},{\bf r}\rightarrow \infty)=\left|\left\langle e^{i{\bf
      k}\cdot {\bf D}} \right\rangle\right|^2=e^{-\frac{1}{3}k^2\sigma_D^2+\cdots}\ .
\ee
Here, $\sigma_D^2\equiv \langle D^2\rangle$. The neglected terms
$\cdots$ in the last expression are non-Gaussian terms in the ${\bf
  D}$ field. 

The ${\bf D}$ field is determined by the particle distribution. So
both the Poisson fluctuation and intrinsic fluctuation in the particle
distribution contribute.  Poisson fluctuation is completely fixed by
the mean number density $\bar{n}_P$ ($f$).  It generates (paper I)
\be
\label{eqn:Poisson}
\sigma_D^2\xrightarrow{\rm Poisson\  limit}0.347L_P^2\ .
\ee
Here $L_P$ is the mean separation of 
particles.  The intrinsic clustering further increases $\sigma_D$. 

If Poisson fluctuation in the particle number
distribution dominates over the intrinsic LSS fluctuation and if the
Gaussian term dominates in Eq. \ref{eqn:SKP}, we predict
$S(k=0.1\hmpc)=0.853$ and $\eta(k=0.1\hmpc)|f)=0.858$ for  $f=0.1\%$ ($\bar{n}_P=6.2\times 10^{-4}
(\mpch)^{-3}$).  meaning $15\%$ systematic underestimation of the velocity power
spectrum. This prediction is already in very good agreement with
numerical result ($\eta=0.87$, Fig. 4). More accurate prediction
requires numerical evaluation of the ${\bf D}$ field statistics in
\S \ref{subsec:D}.

\bfi{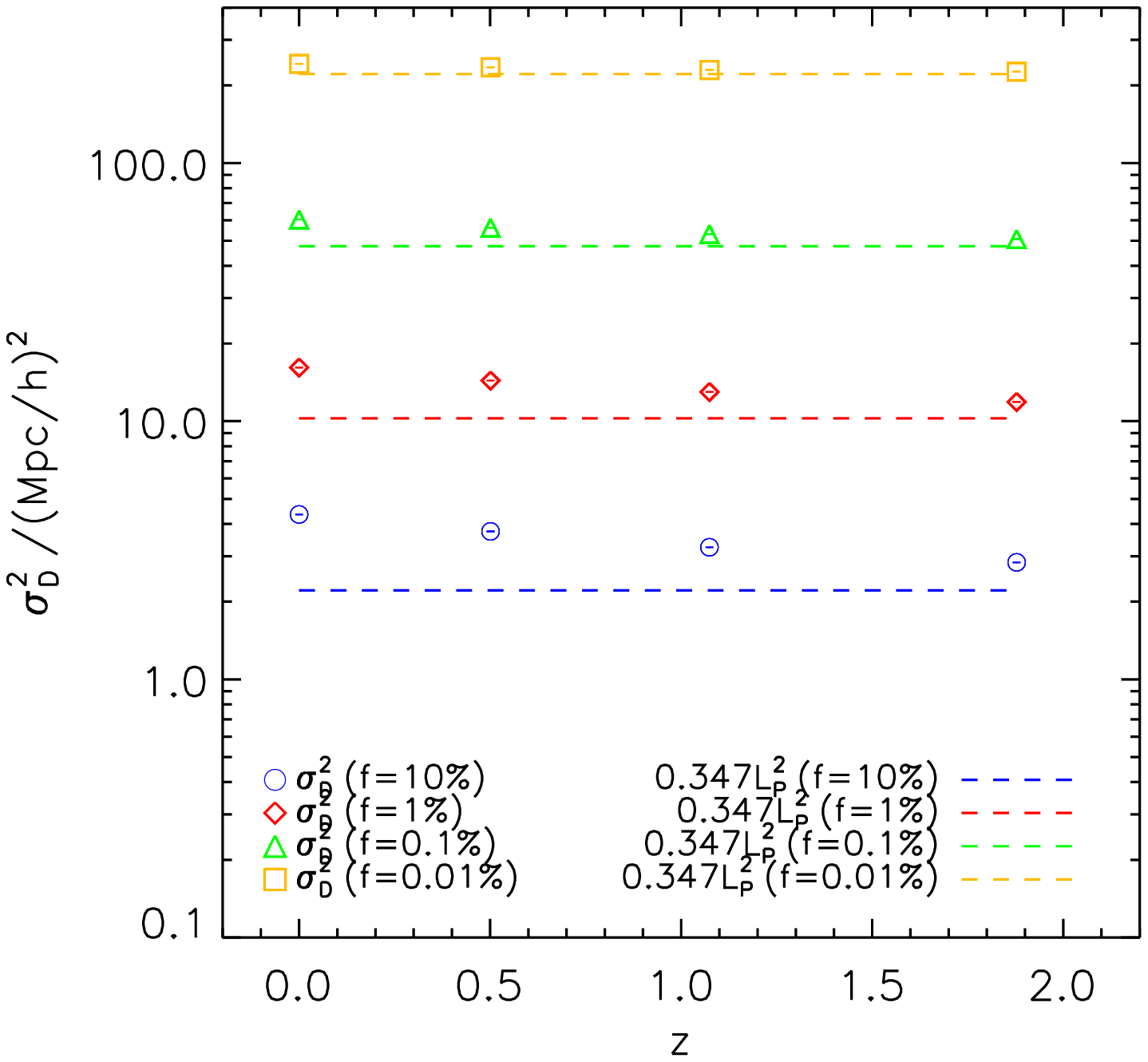}
\caption{$\sigma_D^2\equiv \langle D^2\rangle$ measured in the
  simulation. It is the most important parameter
  determining the sampling artifact. It has two contributions, the
  Poisson fluctuation and the intrinsic LSS fluctuation in the
  particle distribution. A pure Poisson distribution predicts
  $\sigma_D^2=0.347L_P^2$ (dash lines).  For $f\ga 1\%$ ($\bar{n}_P\ga 6.2\times
  10^{-3}(\mpch)^{-3}$), the intrinsic LSS fluctuation significantly
  increases $\sigma_D$ over the Poisson 
  limit. For smaller $f$ (lower number density) and  higher redshift,
  Poisson fluctuation dominates.  \label{fig:sigmaD}}
\efi
\subsection{Statistics of the ${\bf D}$ field}
\label{subsec:D}
The ${\bf D}$ field is the key ingredient to understand the sampling
artifact. In simulations, we can directly measure this field. Relevant
statistics that we measure are  (1) $\sigma_D\equiv \langle D^2\rangle$, (2) the
non-Gaussian measures including  the reduced kurtosis $K_4$ and the 6-th
order cumulants $K_6$ of $D_x$ (equivalently $D_y$ and $D_z$), and (3) the two-point
correlation function of ${\bf D}$. 

\bfi{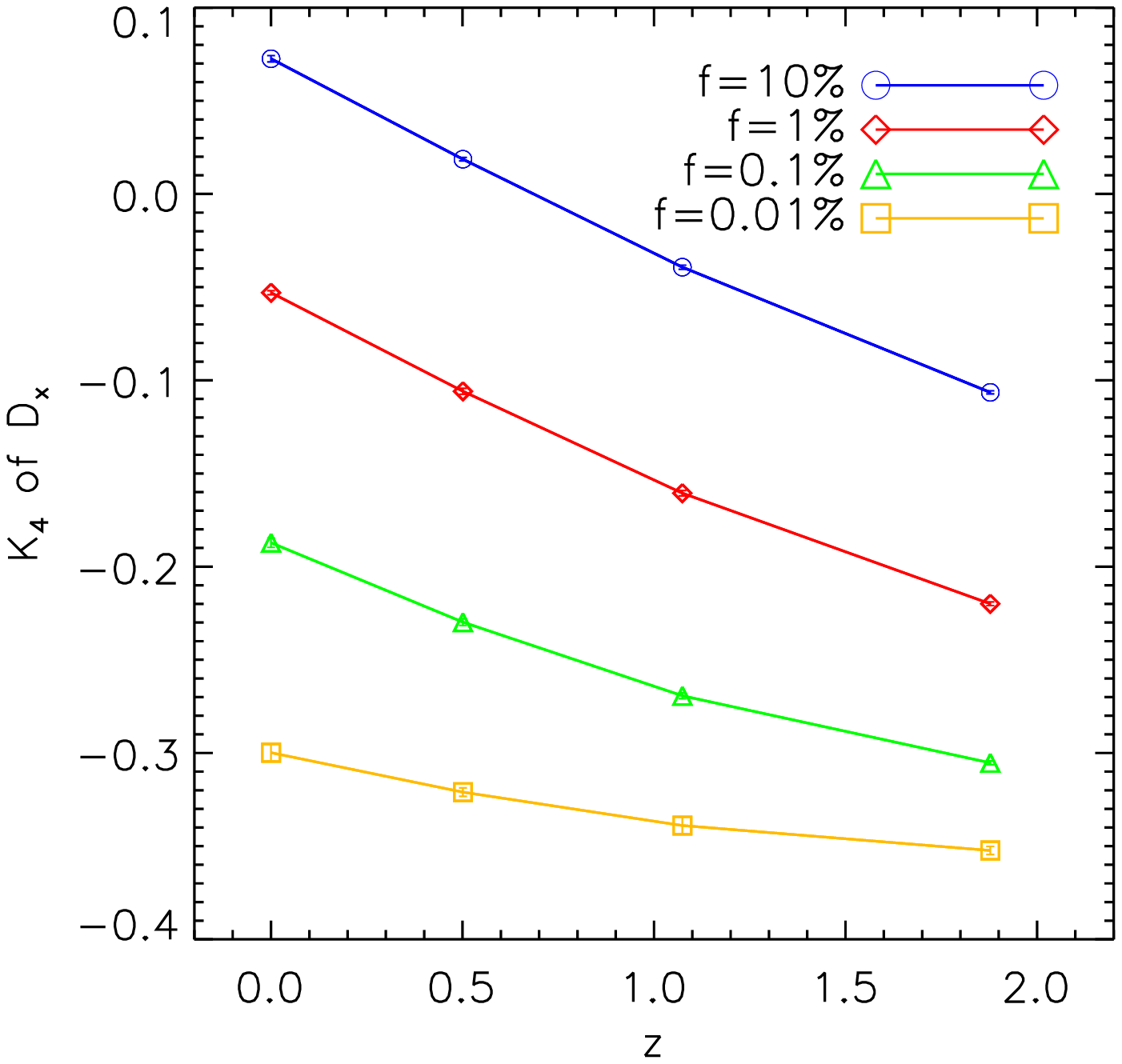}
\caption{The reduced kurtosis $K_4\equiv \langle D_x^4\rangle/\langle
  D_x^2\rangle-3$. Due to the symmetric distribution in $D_x$, this is
the lowest order measure of non-Gaussianity in the ${\bf D}$
field. For practically all relevant cases, the non-Gaussianity
measured by $K_4$ is small. Nevertheless, it is non-negligible in
modelling of the sampling artifact. \label{fig:K4}}
\efi

\subsubsection{The r.m.s dispersion of the ${\bf D}$ field}
$\sigma_D$ governs the overall amplitude of the sampling artifact. The
larger $\sigma_D$, the stronger the suppression to the velocity power
spectrum (Eq. \ref{eqn:SKP}).  Result on $\sigma_D$ at $z\in[0.0,2.0]$
is shown in Fig. 6. As a reminder, both the Poisson fluctuation and
the intrinsic LSS fluctuation in the particle distribution affect
$\sigma_D$. 
(1) We find that the Poisson
approximation (Eq. \ref{eqn:Poisson}) is excellent for $f<0.1\%$
($\bar{n}_P<6.2\times 10^{-4} 
(\mpch)^{-3}$). But for $f=0.1\%$ we begin to observe visible
deviation at $z\la 0.5$. For
$f=10\%$ ($\bar{n}_P=0.062 (\mpch)^{-3}$), $\sigma_D^2$ at $z=0$ is twice of the 
Poisson limit. So the contribution from intrinsic LSS fluctuation is significant. (2)
$\sigma_D$ increases when redshift decreases. Again it manifests the
role of intrinsic LSS fluctuation. It enlarges
$\sigma_D$. When it grows with  decreasing redshift, it causes $\sigma_D$ to increase.

To understand the competition between the Poisson
fluctuation and intrinsic LSS fluctuation, we
estimate the r.m.s density fluctuation generated by the two over the grid size $L_{\rm
  grid}=4\mpch$. For Poisson fluctuation, it is
$\delta_N=(4^3f)^{-1/2}=3.95(f/0.1\%)^{-1/2}$. The intrinsic linear LSS density
fluctuation is $\delta_I\sim 1.7 D(z)$ at $z\gg 1$ at such scale. So it is 
subdominant to $\delta_N$. Here $D(z)$ is the linear density growth
factor. However, due to the faster
growth caused by the nonlinear evolution, $\delta_I(z=0)\simeq
\delta_P(f=0.1\%)$. We can then draw a general conclusion that none of
them overwhelms the other for $f\sim 0.1\%$. It is for
this reason $\sigma_D$ shows visible redshift evolution for
$f=0.1\%$. It also explains why the redshift evolution becomes
significant for $f\ga 1\%$. 

We also find that the contribution from the Poisson fluctuation to
$\sigma_D$ is larger than its contribution to the overall  fluctuation
in the particle number distribution. For example, when $f=100\%$,
$\delta_N\ll \delta_I$ at all relevant redshifts. Nevertheless, we
still find significant contribution to $\sigma_D$ from the Poisson
fluctuation. The Poisson fluctuation scales as
$\delta_N\propto k^2$. The intrinsic LSS fluctuation scales as
$\delta_I\propto k^{(n_{\rm eff}+3)/2}$. When $k\gg k_{\rm NL}$,
$n_{\rm eff}\rightarrow -3$. Hence towards smaller scales, Poisson
fluctuation increases with respect to the intrinsic LSS  fluctuation. We speculate that the ${\bf D}$ field is more sensitive to
smaller scale density fluctuations.

\bfi{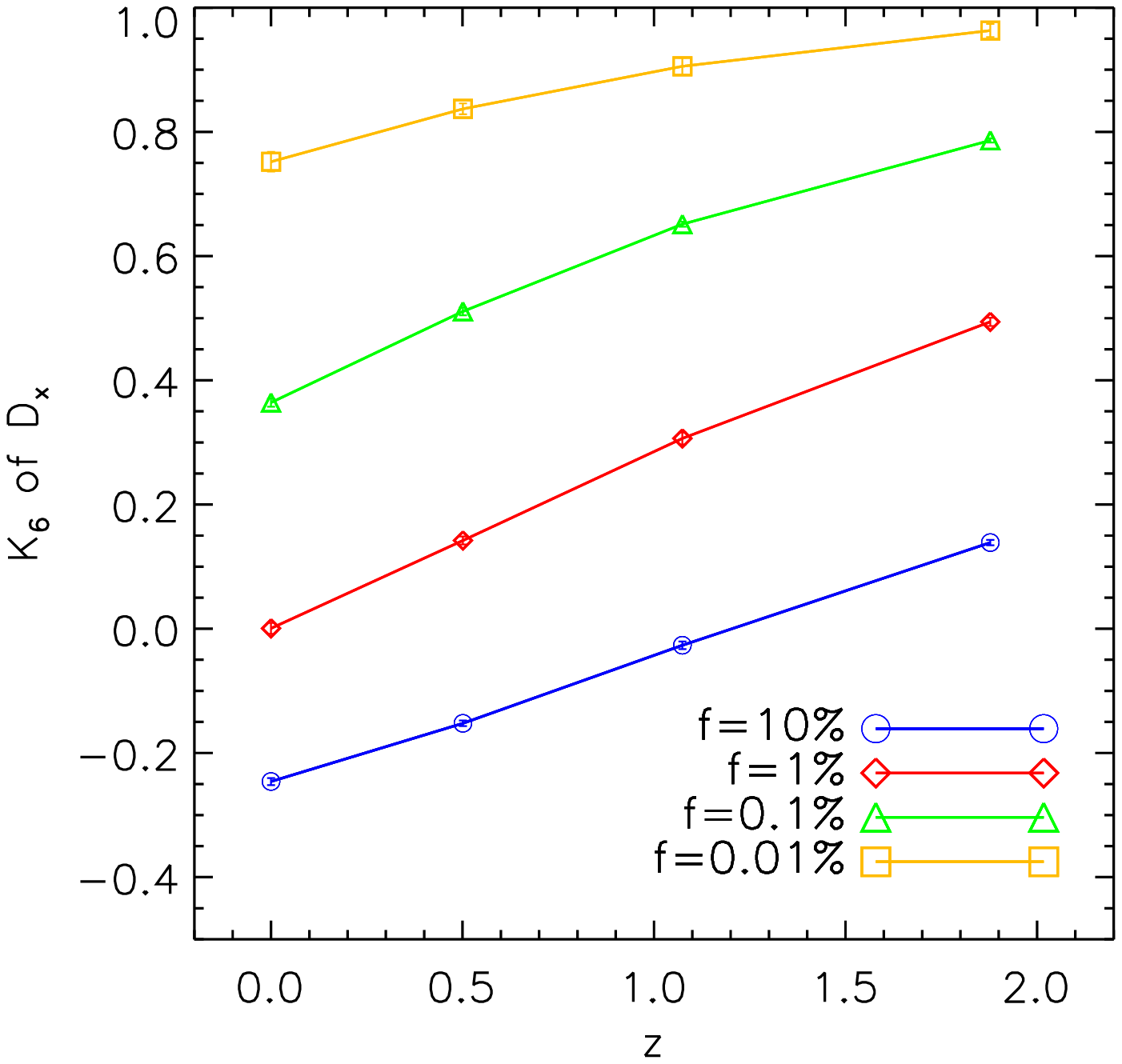}
\caption{The reduced 6-th order cumulant $K_6$. The measured
  non-Gaussianity is insignificant. However, it is still
  non-negligible in modelling the sampling artifact. \label{fig:K6}}
\efi

\subsubsection{Non-Gaussianities in the ${\bf D}$ field}
Eq. \ref{eqn:SKP} tells us that non-Gaussian terms also contribute to $S(k)$ and
hence to the sampling artifact. For
this reason we also measure the reduced 4-th and 6-th order cumulants
for $D_x$  (Fig. 7 \&  8). As a reminder,  $K_4\equiv \langle
D_x^4\rangle/\langle D_x^2\rangle^2-3$ and $K_6=\langle D_x^6\rangle/\langle
D_x^2\rangle^3-15\langle D_x^4\rangle/\langle D_x^2\rangle^2+30$. We do not find very 
significant non-Gaussianities. Nevertheless, the detected non-Gaussianity is
 not negligible. Hence in calculating $S(k)$,
in general we should not use the 
Gaussian approximation in Eq. \ref{eqn:SKP}. Instead, we should directly use the
definition $S(k)\equiv \langle \exp(i{\bf k}\cdot{\bf D})\rangle^2$ to
calculate $S(k)$, since ${\bf D}$ is known in simulation or analysis
of galaxy velocity data. 

\bfi{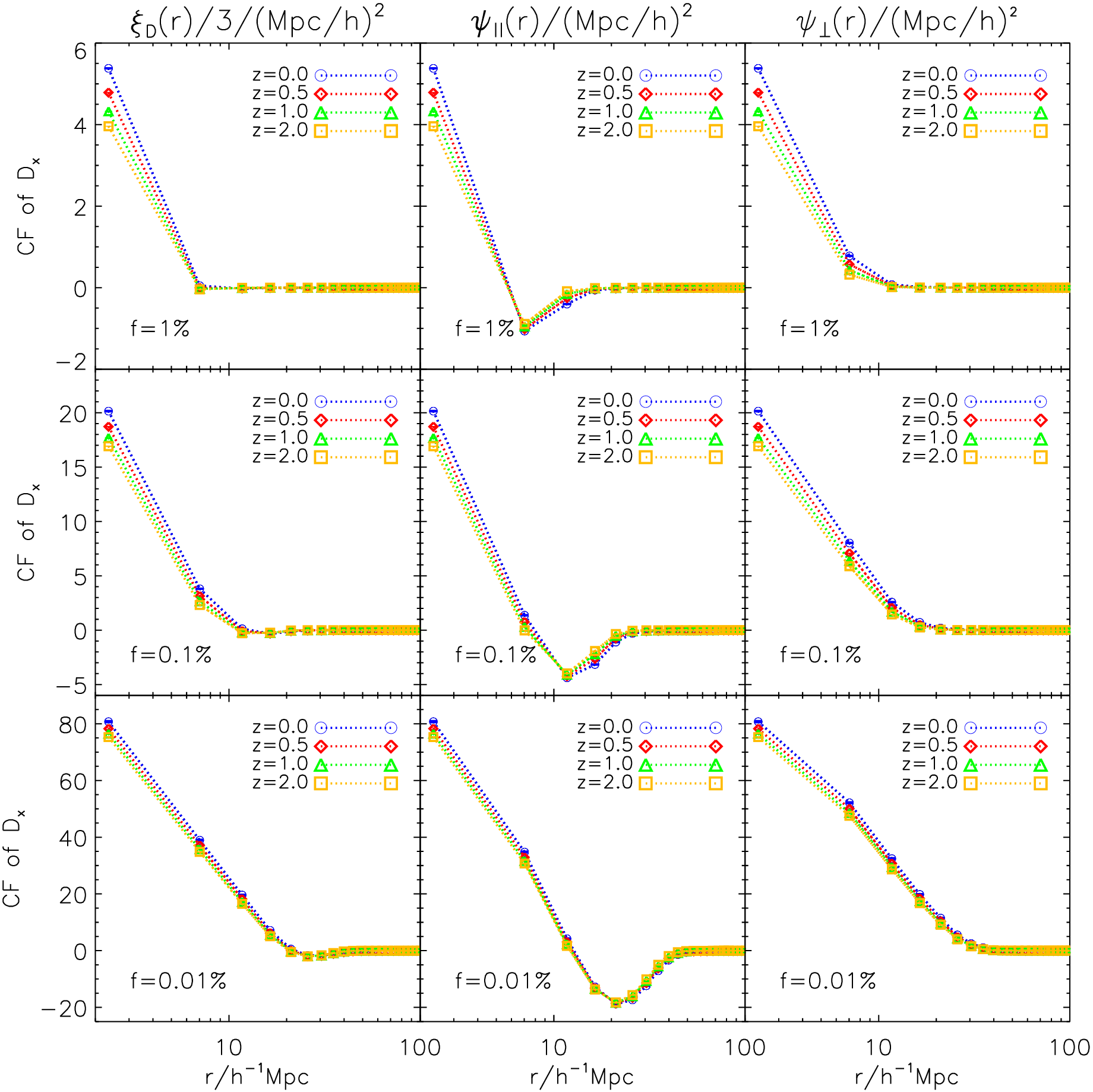}
\caption{Spatial clustering in the ${\bf D}$ field. We show $\xi_D(r)/3=\langle D_xD_x\rangle=(2\psi_\perp+\psi_\parallel)/3$, $\psi_\parallel$ and $\psi_\perp$
  respectively in left, middle and right panels.  The ${\bf D}$ field is
  spatially clustered at $r\la L_P^{-1}$. The mean particle
  separation $L_P=12(f/0.1\%)^{-1/3} \mpch$ for the specific J1200 simulation. \label{fig:xi}}
\efi

\subsubsection{Spatial correlation in the ${\bf D}$ field}
The ${\bf D}$ field is spatially correlated. The spatial correlation
can arise from Poisson fluctuation. This is a little bit surprising
since Poisson fluctuation is not spatially correlated. The reason is
that, for sparse samples,   a significant fraction of particles can be
assigned to more than one grid point and hence build spatial
correlation over scales $\sim L_P$. Intrinsic LSS fluctuation creates
larger voids, in which spatial 
correlation over larger separation can be built. More discussion on
this issue can be found in paper I. 

Following \cite{Peebles80}, we
decompose the correlation function into a perpendicular part
$\psi_\perp$ and a parallel part $\psi_\parallel$,
\be
\langle D_i({\bf x})D_j({\bf x}+{\bf r})\rangle_{\bf
  x}=\psi_\perp(r)\delta_{ij}+(\psi_\parallel(r)-\psi_\perp(r))\hat{r}_i\hat{r}_j\ .
\ee
Here $i,j=x,y,z$ are three Cartesian axes. 
The averaged correlation function
\be
\xi_{\rm D}\equiv \langle {\bf D}({\bf x})\cdot {\bf D}({\bf x}+{\bf
  r})\rangle_{\bf x}=2\psi_\perp(r)+\psi_\parallel(r)\ .
\ee

Fig. \ref{fig:xi} shows the correlation function $\psi_\perp$, $\psi_\parallel$
and $\xi_D/3$, for $f=1\%$, $0.1\%$ and $0.01\%$. As a
reminder, the mean simulation particle separation for the J1200 simulation is
$L_P=12\mpch (0.1\%/f)^{1/3}$.  Indeed, we find non-negligible correlation
at $r\la L_P$. When $r> L_P$, the correlation quickly vanishes and the
${\bf D}$ field can then be treated as a random field of no spatial
correlation. This means that to model the sampling artifact at $k\ll
1/L_P$, we can treat the ${\bf D}$ field as uncorrelated. However, at $k\ga
1/L_P$, the spatial correlation in ${\bf D}$ matters. The leading
order approximation for the sampling artifact (Eq. \ref{eqn:LSA})
neglects such spatial correlation, so it loses accuracy at $k\ga
1/L_P$. Later we will show that the neglected spatial
correlation can be implemented to improve the model accuracy.  

Similar to the case of $\sigma_D$, both the Poisson fluctuation and the
intrinsic LSS fluctuation in the particle distribution contribute to
$\xi_D$. The former does not vary with redshift, while the latter does.
Hence we can use the redshift dependence of $\xi_D$ to infer the
relative importance of the intrinsic LSS fluctuation. For $f\ga
0.1\%$, we observe this redshift dependence. Especially where
$\xi_D>0$, its strength decreases with increasing redshift. This is
caused by decreasing amplitude of the intrinsic LSS fluctuation.  The
impact of LSS weakens when $f$ ($\bar{n}_P$) decreases and
hence Poisson fluctuation increases. For $f=1\%$, the impact from
intrinsic LSS is very significant.  For $f=0.01\%$, the  impact is barely
visible. For $f=0.1\%$, the impact is neither overwhelming nor
negligible. 

Hence for $\bar{n}_P\ga 6\times 10^{-4} (\mpch)^{-3}$,
modelling the spatial correlation in ${\bf D}$ shall take the
intrinsic LSS fluctuation into account.  This further complicates the
modelling of the sampling artifact. For example, a sample of DM
particles and a sample of halos with the same number density in  the
same cosmic volume in general have different sampling artifacts, due
to  different intrinsic LSS clustering.

\subsection{Testing Eq. \ref{eqn:LSA}}
\label{subsec:test}
Our theory, {\it under the approximation of no spatial
correlation in ${\bf D}$}, predicts through Eq. \ref{eqn:LSA}
\be
\label{eqn:eta1}
\eta(k|f)\simeq \frac{S(k|f)}{S(k|f=1)}\ .
\ee
Since the ${\bf D}$ field is directly measurable in simulations, we
can easily evaluate $S(k)\equiv\langle \exp(i{\bf k}\cdot{\bf D})\rangle^2$
(Eq. \ref{eqn:SKP}) and hence evaluate the above theoretical
prediction. In doing so we have automatically included the effect of intrinsic LSS
fluctuation in the particle distribution. This differs from the
simplified prediction in paper I in which only the Poisson fluctuation
is included.  We compare Eq. \ref{eqn:eta1}  against simulation result  in
Fig. 3, 4 \& 5.  We remind that $S(k)$ is evaluated using
  the exact definition $S(k)\equiv\langle \exp(i{\bf k}\cdot{\bf
    D})\rangle^2$, instead of its Gaussian approximation $\exp(-k^2\sigma_D^2/3)$.  Since $S(k)$ does not depend on the direction of
  $k$, we can choose ${\bf k}=(k,0,0)$. We then have
\ba
 S(k)=\langle \exp(ikD_x)\rangle^2=\exp\left[-y(1-\frac{K_4y}{12}+\frac{K_6y^2}{360}+\cdots)\right]\nonumber
\ea
Here, $y\equiv k^2\langle D_x^2\rangle=k^2\sigma_D^2/3$. Fig. 7 \& 8
show visible non-Gaussianity in the ${\bf D}$ field
($K_{4,6,\cdots}\neq 0$). Including
  these non-Gaussian terms in evaulating $S(k)$ is necessary where $y\ga 1$.  For this reason, we
  always use the exact definition to evaluate $S(k)$, instead of using
  the approximation $S(k)\sim \exp(-y)$.

Eq. \ref{eqn:eta1}  shows good to excellent agreement with simulation
results. It well reproduces the overall behavior of increasing
$1-\eta$ with decreasing $f$ ($\bar{n}_P$) and increasing $k$.
Furthermore,  for $f\ga 1\%$ ($\bar{n}_P\ga 6.2\times 10^{-3}
(\mpch)^{-3}$), it is accurate to $\sim 1\%$ or better  over
practically all scales at $k<0.3\hmpc$. For lower number density, the
agreement is worse. Nevertheless, it is still reasonably good. For
example, the theory predicts  $\eta(k=0.1\hmpc|f=0.1\%)=0.82$ at
$z=0$,  compared to the simulation result  $\eta(k=0.1\hmpc|f=0.1\%)=0.87$.

\subsection{More accurate ansatz to model the sampling artifact}
Agreement at such level is encouraging, however not sufficient if we
want to measure the velocity bias of $10^{13}M_\odot$ in simulation to $1\%$ level
accuracy.  These halos have $\bar{n}_P\sim 10^{-3} (\mpch)^{-3}$ at
$z=0$. The $1\%$ accuracy is required to match the stage IV dark 
energy surveys such as BigBOSS/MS-DESI \cite{BigBOSS}, Euclid and
SKA. To achieve this accuracy,  the sampling 
artifact should be corrected to $1\%$ at least at $k=0.1\hmpc$. Eq. \ref{eqn:LSA} is only
able to do so with $\sim 6\%$ accuracy at $k=0.1\hmpc$ for $\bar{n}_P=6\times 10^{-4}
(\mpch)^{-3}$. So further improvement is needed. 

A major source of inaccuracy of Eq. \ref{eqn:LSA}
(Eq. \ref{eqn:eta1}) is the neglected
spatial correlation in ${\bf D}$ when deriving it. {\bf Fig. 9 shows that
the spatial correlation in ${\bf D}$ is non-negligible when spatial
separation is $\la L_P$. Hence it must be incorporated appropriately
in the modelling. } Paper I derives
analytical expression for these corrections.  It is mathematically
solid. Unfortunately it is computationally expensive and is hence hard
to implement in numerical 
evaluation. 

Therefore here we propose an approximate but efficient way  to
incorporate the neglected spatial correlation of ${\bf D}$ into
account, with the hope to improve
over Eq. \ref{eqn:LSA} (Eq. \ref{eqn:eta1}). Using the cumulant expansion theory, Eq. \ref{eqn:S} \& \ref{eqn:SKP} read
\be
S({\bf k},{\bf r})=S(k)e^{k_ik_j\langle D_iD_j\rangle+\cdots}\ .
\ee
Neglecting all high order terms and approximating $\langle
D_iD_j\rangle$ with the one averaged over all directions,  we obtain
\be
S({\bf k},{\bf r}) \simeq
S(k)e^{\frac{1}{3}k^2\xi_D(r)}\ .
\ee
Eq. \ref{eqn:vv2}, \ref{eqn:W} \& \ref{eqn:S} suggest that the
dominant suppression to $\hat{P}(k)$ comes from $S({\bf k},{\bf r})$
with $r\sim 1/k$. Let us approximate that it comes from a single $r_{\rm
  eff}=\alpha/k$, where $\alpha\sim 1$ is a unknown constant to be fixed.  We then expect 
\ba
\label{eqn:ISA}
\hat{P}_E(k)&\simeq &P_E(k) S(k)e^{\frac{1}{3}k^2\xi_D(r_{\rm eff}=\alpha/k)}\\
&=&P_E(k) \langle
e^{i{\bf k}\cdot{\bf D}}\rangle^2e^{\frac{1}{3}k^2\xi_D(r_{\rm eff}=\alpha/k)}\no \ .
\ea
The derivation is far from strict so we shall only treat the
  above result as an approximate ansatz. Nevertheless, it is physically motivated, convenient to
implement and takes the leading order effect of spatial
correlation in ${\bf D}$ into account. (1)  It has the correct asymptotic behavior at
$k\rightarrow 0$,  where the correction vanishes and one recovers the no
spatial clustering limit. (2)  It has the correct asymptotic behavior
at $\alpha\gg 1$. This corresponds to the case of no spatial
clustering (Eq. \ref{eqn:LSA}). (3) For $k=0.1\hmpc$, correlation
at $r=1/k=10\mpch$ is non-negligible for $f<0.1\%$. Furthermore,
$\xi_D>0$ there. So the above formula predicts larger $\eta$ and
hence better agreement with simulation result. 

We do not attempt to find the best-fit $\alpha$.   Instead, we demonstrate the improvement over Eq. \ref{eqn:LSA} with
$\alpha=1/2$ in Fig. \ref{fig:eta1}, 
\ref{fig:eta01} \& \ref{fig:eta001}. The improvement is
significant. For example, it improves the theory accuracy of
$\eta(k|f=0.1\%)$ at $k=0.1\hmpc$ from $\sim 6\%$ to $1\%$. The
improvement for $f=0.01\%$ is even more significant.

\section{Self-calibration and discussion}
\label{sec:conclusion}
The ultimate goal of the current paper and paper I is to correct for the sampling
artifact robustly in order to measure the {\it volume weighted} halo velocity power spectrum
and halo velocity bias accurately. The sampling artifact is completely
determined by $\bar{n}_P$ and the intrinsic LSS fluctuation. Based on
general argument on the two factors, we 
obtain a quick-to-implement ansatz (Eq. \ref{eqn:ISA}) on how the sampling artifact
suppresses the measured velocity power spectrum. 

We have demonstrated that it works  for a variety
of DM samples with the mean particle number density over 4 decades
($6\times 10^{-1}(\mpch)^{-3}$-$6\times 10^{-5}(\mpch)^{-3}$), and 
typical intrinsic LSS clustering from $z=2$ to $z=0$. The
derivation on Eq. \ref{eqn:ISA} is general in the sense that it
assumes no special form of intrinsic LSS fluctuation. Hence as long as it works for DM particles, it should
work equally well for DM halos. With this reasonable extrapolation,
we believe the following self-calibration works for DM halos,
\be
\hat{P}_E(k)\rightarrow \frac{\hat{P}_E(k)}{\langle
e^{i{\bf k}\cdot{\bf D}}\rangle^2e^{\frac{1}{3}k^2\xi_D(r_{\rm
    eff}=\alpha/k)}} .
\ee
We caution that now the ${\bf D}$ field is that of DM halos, which
differs from that of DM particles. 
This measure of the velocity power spectrum at $k\leq 0.1\hmpc$ should
be essentially free  of the sampling artifact, at the level of $1\%$ for $10^{13}M_\odot$
halos or less massive ones at $z=0$. This will then allow us to
measure the {\it real} halo velocity 
bias, free of otherwise severe systematic error from the sampling
artifact. We will present such measurements in \cite{Zheng14b}, which
belongs to our ongoing efforts to understand the velocity field, redshift
shift space distortion and  velocity reconstruction in spectroscopic
redshift surveys \cite{paperI,paperII}.  

The current paper focuses on the sampling artifact in the
  gradient part of the velocity field. The curl part
  of the velocity also suffers from numerical artifacts such as the
  alias effect \cite{Pueblas09} and the sampling artifact
  \cite{Zhang14,paperII}.  We can use the same method
  (Eq. \ref{eqn:eta}) to quantify the sampling artifact in the curl
  part of the velocity field. One subtlety is that numerical artifacts
in the curl velocity are much more severe, so  higher simulation
resolution is required than the J1200 that we have analyzed.  This
topic will be explored elsewhere. 

\section{Acknowledgement}

This work was supported by the National Science Foundation of China (Grants No. 11025316, No. 11121062, No. 11033006, No. 11320101002, and No. 11433001), National Basic Research Program of China (973 Program 2015CB857000), the NAOC-Templeton Beyond the Horizon program, the CAS Strategic Priority Research Program "The Emergence of Cosmological Structures" of the Chinese Academy of Sciences, Grant No. XDB09000000 and the key laboratory grant from the Office of Science and Technology, Shanghai Municipal Government (No. 11DZ2260700). P. J. Z. gratefully acknowledges the support of the National Science Foundation through Grant No. PHYS-1066293 and thanks the Simons Foundation and, for their hospitality, the Aspen Center for Physics, where part of the work was done. Y. Z. thanks Yu Yu, Yanchuan Cai, and Jiawei Shao for the useful discussions.

\appendix

\bfi{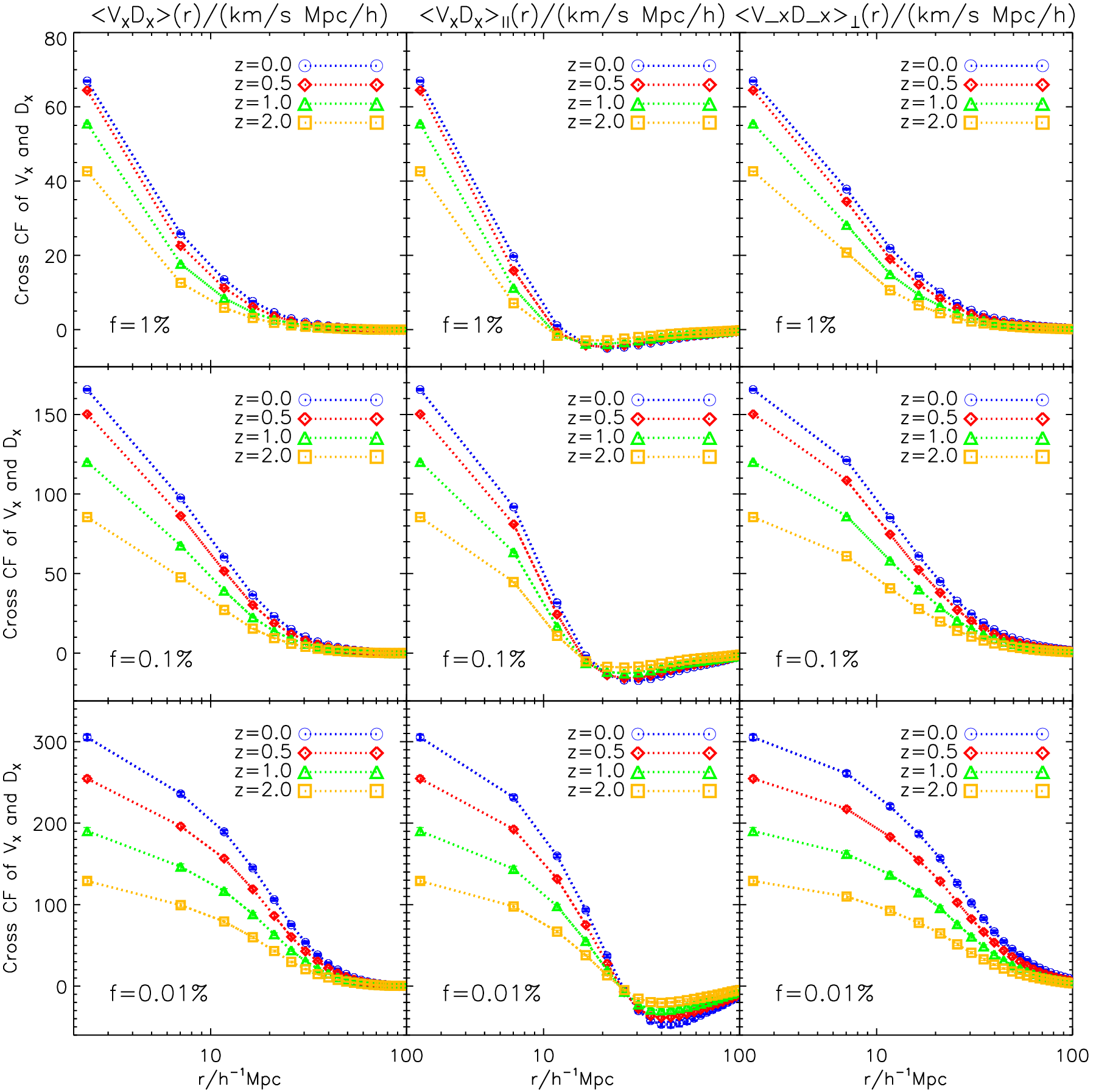}
\caption{The  ${\bf v}$-${\bf D}$ correlation detected in
  simulations. The left, middle and right panels are $\langle
  v_xD_x\rangle=(2\langle vD\rangle_\perp+\langle
  vD\rangle_\parallel)/3$, $\langle
  vD\rangle_\parallel$ and $\langle vD\rangle_\perp$
  respectively. The ${\bf v}$-${\bf D}$ correlation is weak in that
  $\langle vD\rangle\ll \sigma_v\sigma_D$. However,  it is in
  principle a significant aspect of  the sampling artifact and
  should be taken into account, especially at $k>0.1\hmpc$ for sparse samples. \label{fig:vD}}
\efi

\bfi{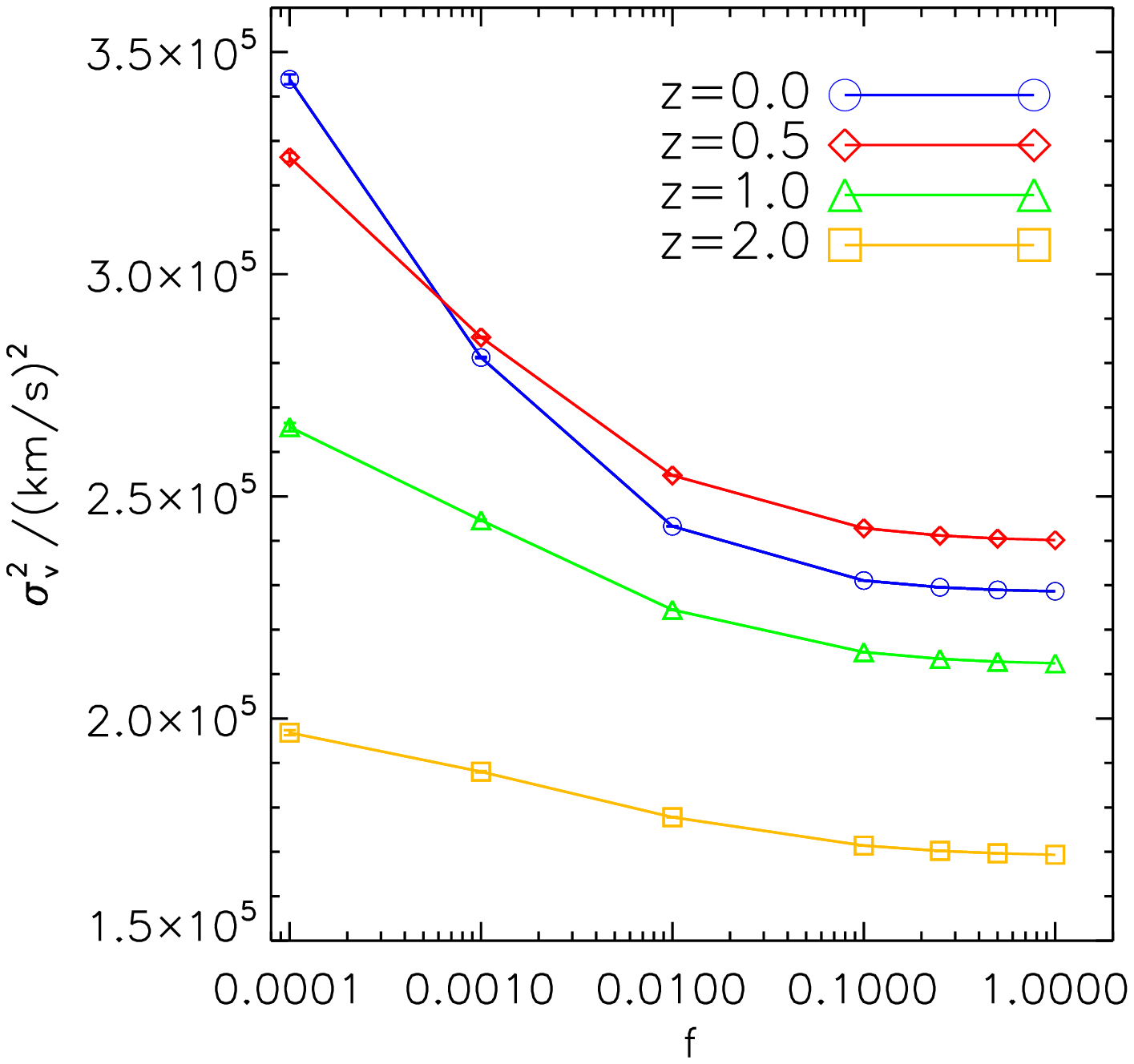}
\caption{$\sigma_v^2\equiv \langle \hat{v}^2\rangle$ as a function of
$\bar{n}_P=0.62f(\mpch)^{-3}$ and
$z$. Its variation with $f$ is caused by correlation between the
velocity (signal) field and the ${\bf D}$ (sampling) field (the appendix). It 
highlights another aspect of the sampling artifact. In 
contrast to underestimation of  the velocity power spectrum at large
scale, the sampling artifact overestimates the velocity dispersion.
It is likely responsible for the overestimation of the velocity power
spectrum at small scales and low $f$, observed in
Fig. \ref{fig:eta001} for the case of $f=0.01\%$. \label{fig:v2}}
\efi

\section{More aspects of the sampling artifact}
\label{sec:NA}

The $\eta$ measurement for $f=0.01\%$ (Fig. \ref{fig:eta001}) shows
abnormal behaviors at $k\ga 0.08\hmpc$. The most significant is
the turn-over at $k\sim 0.08 \hmpc$ and the eventual $\eta>1$ at $k\ga
0.15 \hmpc$, for $z=0$. Another anomaly is that $\eta$ decreases with
increasing redshift, in contrast to our theoretical expectation and
the cases of $f\ga 0.1\%$.  The two anomalies are likely related. These
anomalies are not statistical flukes, since  we have run many more
realizations of DM sub-samples and found the same anomalies. They may
imply either unknown numerical artifacts or inappropriate understanding
of the sampling artifact in very sparse samples. 

Unfortunately so far we do not have any real insight to solve these
issues. We are only able to discuss/evaluate some possibilities, and
we caution the readers that this list is likely not exhaustive.   
\bi
\item Transport of power of ${\bf v}$ across scales by the field ${\bf
    D}$. This is caused by spatial correlation of the ${\bf D}$ field,
exactly analogous to the deflection field in CMB 
lensing \cite{Seljak96}. Where the real signal is weak,
we may find overestimation of the velocity power
spectrum. This point can be demonstrated by a toy model, in which
$P_{ij}({\bf k})=A_{ij}$ if ${\bf k}={\bf k}_*$ and zero
otherwise. $\hat{P}_{ij}({\bf k}_*)=A_{ij}W({\bf k}_*,{\bf
  0})=A_{ij}S(k_*)<A_{ij}=P({\bf k}_*)$. The lost power is
transported to other modes,  $\hat{P}_{ij}({\bf k}\neq {\bf k}_*)=A_{ij} W({\bf k}_*,{\bf
  k}_*-{\bf k})\neq 0$. Is it sufficient
to explain the observed anomalies in Fig. \ref{fig:eta001}? We notice
that the clustering strength of the ${\bf D}$ field changes little
between $z=0$ and $z=2$ (Fig. \ref{fig:xi}), while $\eta$ increases dramatically at
$k=0.1\hmpc$ from $z=2$ to $z=0$. This implies that, the transport of
power of ${\bf v}$ by the ${\bf D}$ field is not the major cause of
the observed anomalies in Fig. \ref{fig:eta001}. 
\item A more likely cause is the ${\bf v}$-${\bf D}$ correlation, neglected
in the theoretical modelling. It is distinctively different to CMB
lensing, in which the lensing field and primary CMB have no cross-correlation. As discussed in paper I, the ${\bf
  D}$ field is spatially correlated to the velocity field. It can not
only transport power across scales, but also generate extra power in
${\bf v}$.  This correlation is neglected in
Eq. \ref{eqn:vv2} and all results derived based on Eq. \ref{eqn:vv2}
(refer to more details in paper I). 
\ei

\section{The detected ${\bf v}$-${\bf D}$ correlation and
  its impact on the sampling artifact }
\label{sec:vD}
The ${\bf v}$-${\bf D}$ correlation is inevitable since the intrinsic LSS 
fluctuation, a source of ${\bf D}$,  correlates with ${\bf
  v}$. We confirm its existence by simulations
(Fig. \ref{fig:vD}). Like the case of auto correlation in ${\bf D}$,
the cross correlation can also be decomposed into two coordinate
independent components,
\be
\langle v_i({\bf x})D_j({\bf x}+{\bf r})\rangle=\langle
vD\rangle_{\perp}\delta_{ij}+(\langle vD\rangle_{\parallel}-\langle
vD\rangle_{\perp})\hat{r}_i\hat{r}_j\ .
\ee
The measured cross correlation is weak, comparing to the auto
power. For example, for $f=0.1\%$,  $\sigma_D\sigma_v\simeq 2400$ km/$s$
Mpc$/h$.  So $\langle vD\rangle\ll \sigma_v\sigma_D/3$.  This is
expected, since only the part of ${\bf D}$ sourced by the intrinsic LSS fluctuation is
correlated with ${\bf v}$. 

We have neglected this complexity of ${\bf v}$-${\bf D}$ correlation
in modelling the sampling artifact.   This is a
major drawback of our theoretical modelling. In particular,  it could be the major
course of the observed anomalies in Fig. \ref{fig:eta001} at $k\ga
0.1\hmpc$. It may also be partly responsible for the discrepancy
between the simplified theoretical modelling and simulation results in
Fig. \ref{fig:eta01}.  Unfortunately, we are not able to implement it into
quantitative theoretical calculation yet. Therefore we are not able to
directly testify the above speculations. Nevertheless, we can prove
that it indeed has significant impact on a highly related statistics, $\langle \hat{v}^2\rangle$. 

If this correlation is indeed
negligible, we prove in \S \ref{sec:v2} a unbiased velocity dispersion
measurement, $\langle 
\hat{v}^2\rangle=\langle v^2\rangle$. Hence $\langle 
\hat{v}^2\rangle$ should be  independent of the particle fraction $f$. 
However, simulations show that $\langle\hat{v}^2\rangle$ increases
with decreasing $f$ (Fig. \ref{fig:v2}). It clearly
proves the significance of correlation between ${\bf v}$ and
${\bf D}$.   It causes the velocity
dispersion to be overestimated.  For
$f=0.01\%$, the overestimation reaches $\sim 20\%$ at $z=2$ and $\sim
50\%$ at $z=0$.  Since $\langle v^2\rangle$ is the integral of the power spectrum,
 overestimation in $\langle v^2\rangle$ must  also show up as overestimation of
 the power spectrum at certain scales. Hence it should be
 responsible for  the observed anomalies in Fig. \ref{fig:eta001}. 

This overestimation of $\langle v^2\rangle$  is a new impact of the
sampling artifact. It arises from the fact that the weighting assigned
to each particle is correlated with the velocity (signal) field.  On
the average, the weighting of each particle in the volume weighted
scheme is $\propto (1+\delta)^{-1}$. $\delta$
is the combination of the underlying DM density 
fluctuation and Poisson fluctuation.   The Poisson fluctuation is
uncorrelated with the velocity field. However, the intrinsic fluctuation is {\it
  positively} correlated with the local velocity dispersion, resulting
in a positive correlation between $\delta$ and the local velocity
dispersion. The weighting $\propto (1+\delta)^{-1}$ then suppresses contribution of high
density/high velocity dispersion regions. Sparser samples have 
larger Poisson fluctuation and hence weaker correlation between the simulated
$\delta$ and the local velocity dispersion. Therefore it suffers from weaker
suppression of high {\it real} density/high velocity dispersion
regions. So decreasing the particle number density increases
$\langle \hat{v}^2\rangle$. 

The same intrinsic LSS fluctuation causing
correlation between the weghting and the velocity signal also causes
${\bf v}$-${\bf D}$ correlation. So the two explanations are
consistent.

\section{$\langle
\hat{v}^2\rangle=\langle v^2\rangle$ if ${\bf v}$ and ${\bf D}$ are
uncorrelated}
\label{sec:v2}
For brevity, we work at the limit of infinite box size and
infinitesimal grid size. The proof for finite box size and non-zero
grid size is similar. 
\ba
\langle \hat{v}^2\rangle&=&\int \hat{v}^2({\bf x})\frac{d^3{\bf x}}{V}\\
&=&
\int \langle {\bf v}({\bf k})\cdot {\bf v}({\bf q})e^{i({\bf k}+{\bf q})\cdot
  {\bf D}}\rangle e^{i({\bf k}+{\bf q})\cdot
  {\bf x}} \frac{d^3{\bf k}d^3{\bf q}d^3{\bf x}}{(2\pi)^6V}\no\ .
\ea
Here $V=\int d^3{\bf x}$ is the total volume. 
When ${\bf v}$ and ${\bf D}$ are uncorrelated, 
\ba
\langle {\bf v}({\bf k})\cdot {\bf v}({\bf q})e^{i({\bf k}+{\bf q})\cdot
  {\bf D}}\rangle\rightarrow \langle {\bf v}({\bf k})\cdot {\bf
  v}({\bf q})\rangle\langle e^{i({\bf k}+{\bf q})\cdot
  {\bf D}}\rangle \no\\
=P_v(k)(2\pi)^3\delta_{3D}({\bf k}+{\bf q})\langle e^{i({\bf k}+{\bf q})\cdot
  {\bf D}}\rangle\no\\
=P_v(k)(2\pi)^3\delta_{3D}({\bf k}+{\bf q})\ .\no
\ea 
We then finally prove
\ba
\langle \hat{v}^2\rangle
=\int P_v(k)\frac{d^3{\bf k}}{(2\pi)^3}
\frac{d^3{\bf x}}{V}=\langle v^2\rangle\ .
\ea
This means that, if the signal (${\bf v}$) and the sampling field
${\bf D}$ are uncorrelated, the estimation of velocity dispersion will
be unbiased. 
A corollary is that, the measured $\langle \hat{v}^2\rangle$ should
not depend on the particle fraction $f$, if ${\bf v}$ and ${\bf D}$ are
uncorrelated. Then if $\langle
\hat{v}^2\rangle$ depends on $f$ ($\bar{n}_P$), there must be non-negligible
correlation between ${\bf v}$ and ${\bf D}$.  The observed significant
dependence of $\langle \hat{v}^2\rangle$ on the particle number
density  (Fig. \ref{fig:v2}) then provides an indirect, nevertheless
solid, evidence  of ${\bf v}$-${\bf D}$ correlation. This is further supported by the direct
measurement in Fig. \ref{fig:vD}.

\bibliography{mybib}

\end{document}